\documentclass[12pt]{article}

\usepackage{amsmath}
\usepackage{amssymb}
\usepackage{amsfonts}
\usepackage{lscape}

\begin{document}

\fontsize{12}{6mm}\selectfont
\setlength{\baselineskip}{2em}

$~$\\[.35in]
\newcommand{\dss}{\displaystyle}
\newcommand{\raro}{\rightarrow}
\newcommand{\be}{\begin{equation}}
\newcommand{\ba}{\end{equation}}

\def\sech{\mbox{\rm sech}}
\def\sn{\mbox{\rm sn}}
\def\dn{\mbox{\rm dn}}
\thispagestyle{empty}

\begin{center}
{\Large\bf Motion on Constant Curvature Spaces } \\ [2mm]   
{\Large\bf and Quantization Using Noether Symmetries}  \\   [2mm]
\end{center}

\vspace{1cm}
\begin{center}
\noindent
{\bf Paul Bracken}            \\            
{\bf Department of Mathematics,}   \\
{\bf University of Texas,}   \\
{\bf Edinburg, TX  }     \\
{78540}
\end{center}

\vspace{2cm}
\begin{abstract}
A general approach is presented for quantizing a metric nonlinear system
on a manifold of constant curvature. It makes use of a curvature dependent
procedure which relies on determining Noether symmetries from the metric.
The curvature of the space functions as a constant parameter. For a specific metric
which defines the manifold, Lie differentiation of the metric gives these 
symmetries. A metric is used such that the resulting Schr\"odinger equation can 
be solved in terms of hypergeometric functions. This permits the investigation
of both the energy spectrum and wave functions exactly for this system.
\end{abstract}

\vspace{1cm}
PACs: 03.65.Ge, 03.65.Ta, 03.65.Aa

\vspace{4mm}

\newpage
\section{Introduction}
\numberwithin{equation}{section}

The standard approach to quantization in quantum mechanics 
usually takes place on a Euclidean space which is characterized
by a zero curvature scalar {\bf [1-2]}. 
The quantization of physical models on a curved space, even a space 
of constant curvature is a problem which impacts many different
areas of physics such as gravitation {\bf [3-4]}.
A specific example of physical
importance is the existence of Landau levels for the motion
of a charged particle under perpendicular fields, which has
been investigated in the case of non-Euclidean geometries {\bf [5]}.
The quantum dot has also given rise to the use of models
which are based in the area of quantum mechanics on spaces
of constant curvature. In fact, the entire area of
gravitation and cosmology is approached at the present time
on a geometric foundation. This relies on specifying a space-time 
characterized by a metric, whose components are used to calculate
the curvature of the spacetime manifold. The result need not always be 
constant, but this case is easier to study mathematically.
Schr\"odinger first made use of a
factorization method for the study of the hydrogen atom in
a spherical geometry {\bf [6]}, and dynamical symmetries in
a spherical geometry have been worked out by Higgs {\bf [7]}. 
A more esoteric problem is the study
of polygonal billiards, or systems which are enclosed by
geodesic arcs on surfaces with curvature. Some motions that 
are integrable in the Euclidean case can become ergodic
when the curvature is negative, so this subject overlaps
with the study of chaos in quantum systems {\bf [8]}.

It is the objective here to look at the quantization of a geometric
model on a curved space which can be thought of as a
two-dimensional oscillator under a particular choice of potential.
An approach is proposed here which ought to be applicable to many
types of models which are specified by a metric
in the sense that a metric is defined and the components of
the metric are used to construct the Lagrangian {\bf [9-11]}. 
The two-dimensional problem, originally introduced
as a nonlinear deformation of a linear system, can in fact be
interpreted as a potential model on a space with constant
curvature. 
The three spaces with constant curvature $\kappa$, the sphere
$S^2_{\kappa} (\kappa >0)$, Euclidean plane $\mathbb E^2$,
and hyperbolic plane $H^2_{\kappa} (\kappa <0)$ can be
thought of as three different cases arising from a
family of Riemannian manifolds $M^2_{\kappa} = (S^2 , 
\mathbb E^2, H^2)_{\kappa}$ with the curvature 
$\kappa \in R$ appearing as a parameter. The components of the
metric will be selected according to this geometry, but 
everything is done in such a way that applications
to other types of systems whose Lagrangian is specified
by a metric should be straightforward.
Other spaces to which the procedure applies would be 
classically diffeomorphic to either a 
sphere of constant curvature, or to the hyperbolic plane 
depending on the sign of the curvature. The curvature is
thus considered as a parameter and all mathematical 
expressions are presented in a curvature dependent way
in terms of this parameter. The ideas of the procedure can be
enunciated in a very
general, mathematical framework than has been done. It is hoped that
these ideas can be applied to other types of metrics to achieve
similar results {\bf [12-14]}. With the metric adopted here, it will be
seen that eigenfunctions can be obtained for the
Schr\"odinger equation. 

The model can be formulated in Cartesian and cylindrical
coordinates and the transformation properties of 
dynamical variables such 
as the Lagrangian under coordinate changes is studied.
The Lagrangian is determined once the components of
the metric have been defined, and depends on the coordinates
of the underlying manifold. It is shown how the Killing vector
fields can be calculated by Lie differentiation of the
metric. These Killing vectors can be shown to be specified by
a coupled system of partial differential equations which can
be solved in closed form for the component functions 
in these vector fields. The Poisson brackets of the classical
variables can be calculated as well as the commutator brackets
of the Killing vector fields. This work can be done quickly
by using symbolic manipulation {[15]}. The Hamiltonian
is calculated by means of the usual canonical transformation.
It is necessary to know the Hamiltonian in order to quantize the system.
The Killing vector fields will provide the Noether momenta
for the system. The canonical momenta do not coincide
with the Noether momenta when $\kappa \neq 0$, and so the quantization procedure
is more complicated in a constant curvature space.
The usual quantization prescription can be applied to
the components of the Noether momenta which appear in
the Hamiltonian. It may be noted that the Poisson momenta
do not Poisson commute, and the corresponding self-adjoint 
quantum counterparts do not commute as operators. Verifying these
statements is possible by using symbolic manipulation again.
To summarize, it is explicitly indicated how the transition
from the classical curvature dependent picture to a quantum
system in terms of operators can be done using this
approach based on the quantization of the Noether momenta.
Finally, it is explained how the exact resolution of the
curvature dependent Schr\"odinger equation can be accomplished.
The eigenfunctions and energies can be calculated and
studied in detail for the metric of this problem, 
and will lead to the concept of
curvature dependent plane waves.

\section{Metric and Lagrangian}

A model will be constructed and examined which is defined by
a specific metric and relates to geodesic motion. The 
dynamics is obtained from a Lagrangian whose kinetic energy
term depends on the curvature parameter of the underlying
space and is related to the metric. Let $M$ be a Riemannian or
pseudo-Riemannian manifold whose metric evaluated at a point 
$p \in M$ is $g(p)$ {\bf [16]}. On the tangent space $T M$, consider
the Lagrangian given by the kinetic energy of the metric
\be
T (v) = \frac{1}{2} \, g_{ij} v^i v^j.
\label{eqII1}
\ba
The general Lagrangian is obtained from \eqref{eqII1} by adding a
potential term. The exact form of the metric $g$ in the case
studied here is defined in cylindrical coordinates to be
\be
g = \frac{1}{2} \frac{1}{1 - \kappa r^2} \, dr \otimes dr + 
\frac{1}{2} r^2 d \varphi \otimes d \varphi.
\label{eqII2}
\ba
In \eqref{eqII2}, $\kappa$ is the curvature scalar and this
metric can be put into an equivalent form in which the
geometry of the manifold is clearer. The three spaces of constant
curvature which occur here, the sphere $S^2_{\kappa}$ $(\kappa >0)$, the
Euclidean plane $\mathbb E^2 (\kappa =0)$, and hyperbolic plane
$H^2_{\kappa}$ $(\kappa <0)$, can be considered
different situations inside a family of Riemannian
manifolds $M^2_{\kappa}= (S^2_{\kappa}, \mathbb E^2, H^2_{\kappa})$
with the curvature $\kappa \in R$ as a parameter.
Taking the components from \eqref{eqII2} and putting them in 
\eqref{eqII1}, the general Lagrangian is given in terms of
cylindrical variables with $v_r = \dot{r}$ and $v_{\varphi} = \dot{\varphi}$ as
\be
L ( \kappa) = \frac{1}{2} ( \frac{v_r^2}{1 - \kappa r^2}
+ r^2 v_{\varphi}^2 ) + V(r).
\label{eqII3}
\ba
It is useful to study this system at the classical level
before proceeding to look at its quantization. 
The standard transformations of Lagrangian mechanics
can be established as well as the transformations
between Cartesian and cylindrical forms. To this end, begin with
the transformation given by
\be
x = r \cos \varphi,
\qquad
y = r \sin \varphi.
\label{eqII4}
\ba
All the variables appearing in \eqref{eqII4} depend on 
an evolution parameter.
Differentiating the variables in the transformation
\eqref{eqII4} with respect to the time parameter $t$, 
setting $v_x = \dot{x}$ and $v_y = \dot{y}$,
these additional relationships are obtained
\be
v_x = v_r \cos \varphi - r \sin \varphi \, v_{\varphi},
\qquad
v_y = v_r \sin \varphi + r \cos \varphi \, v_{\varphi}.
\label{eqII5}
\ba
From \eqref{eqII5}, it follows that
\be
v_x^2 + v_y^2 - \kappa ( x v_y - y v_x )^2 = v_r^2 + r^2 
(1 - \kappa r^2) v_{\varphi}^2.
\label{eqII6}
\ba
The Lagrangian \eqref{eqII3} in Cartesian coordinates is given by
\be
L ( \kappa) = \frac{1}{2} \frac{1}{ 1 - \kappa r^2}
[ v_x^2 + v_y^2 - \kappa ( x v_y - y v_x)^2 ] + V(x,y).
\label{eqII7}
\ba
Using formulas such as \eqref{eqII6}, the Lagrangian can be
transformed from Cartesian to cylindrical form or from 
cylindrical to Cartesian by means of \eqref{eqII6}. 

To obtain a quantum formulation for the system,
it is important to study the Hamiltonian. The Hamiltonian is
determined from the Lagrangian \eqref{eqII7} by first obtaining
the momenta by differentiating $L$, 
\be
p_x = \frac{\partial L}{\partial \dot{x}} = \frac{v_x + \kappa (x v_y - y v_x) y}
{ 1 - \kappa r^2},  \qquad
p_y = \frac{\partial L}{\partial \dot{y}} = \frac{v_y - \kappa ( x v_y -y v_x) x}
{1 - \kappa r^2}.
\label{eqII8}
\ba
Solving \eqref{eqII8} for $v_x$ and $v_y$, we obtain,
\be
v_x = (1- \kappa x^2) p_x - \kappa xy p_y,
\qquad
v_y= (1 - \kappa y^2) p_y - \kappa xy p_x.
\label{eqII9}
\ba
The Hamiltonian is calculated from \eqref{eqII9} and the Lagrangian \eqref{eqII7}
by means of the usual transformation,
\be
H = p_x v_x + p_y v_y - L (\kappa)
= \frac{1}{2} ( p_x^2 + p_y^2 - \kappa (x p_x + y p_y)^2) - V(x,y).
\label{eqII10}
\ba
If \eqref{eqII8} is taken and $p_x$, $p_y$ are substituted back into
$L ( \kappa)$, the Lagrangian \eqref{eqII7} is recovered.
Finally, to obtain the Hamiltonian in terms of the cylindrical $r$, $\varphi$
coordinates, the momenta which correspond to \eqref{eqII3} are
determined by differentiating the Lagrangian in \eqref{eqII3},
\be
p_r = \frac{\partial L}{\partial \dot{r}}
= \frac{v_r}{1 - \kappa r^2},
\qquad
p_{\varphi} = \frac{\partial L}{\partial \dot{\varphi}} = r^2 \dot{\varphi}.
\label{eqII11}
\ba
Therefore, the Hamiltonian in these coordinates is given by
\be
H ( \kappa) = p_r v_r + p_{\varphi} v_{\varphi} - L ( \kappa)
= \frac{1}{2} [ (1- \kappa r^2) p_r^2 + \frac{1}{r^2} p_{\varphi}^2] - V(r).
\label{eqII12}
\ba
Finally, the equations of motion can be written down in the $x$, $y$ 
coordinate system.
With $q=x,y$, the Euler-Lagrange equation will produce these,
\be
\frac{d}{dt} ( \frac{\partial L}{\partial \dot{q}} ) - \frac{\partial L}{\partial q} =0.
\label{eqII13}
\ba
Calculating \eqref{eqII13} for both variables, two coupled equations are
obtained each of which contain both derivatives $\ddot{x}$ and $\ddot{y}$.
Solving this pair as a system in 
the two variables $\{ \ddot{x}, \ddot{y} \}$ with the potential function unspecified
for the moment, the equations of motion are found to be
$$
( 1 - \kappa r^2) \ddot{x} + \kappa ( \dot{x}^2 + \dot{y}^2 - \kappa
( x \dot{y} - y \dot{x})^2 ) x = (1 - \kappa r^2)
( ( 1 - \kappa x^2) \frac{\partial V}{\partial x} - \kappa xy \frac{\partial V}{\partial y}),
$$
$$
( 1- \kappa r^2) \ddot{y} + \kappa ( \dot{x}^2 + \dot{y}^2 - \kappa ( x \dot{y} - y \dot{x})^2)y
= (1- \kappa r^2)( - \kappa xy \frac{\partial V}{\partial x} + ( 1 - \kappa y^2) \frac{\partial V}
{\partial y} ).
$$
The potential that will be used in the subsequent analysis will have the form,
$$
V (r) = - \frac{\alpha^2}{2} ( \frac{r^2}{1 - \kappa r^2}).
$$
In fact, many other forms for the potential are admitted by the method
proposed here. This potential has the advantage that eigenfunctions can
be obtained for its Schr\"odinger equation.
Substituting this potential into the equation of motion, they become
the following
\be
\begin{array}{c}
( 1 - \kappa r^2) \ddot{x} + \kappa [ \dot{x}^2 + \dot{y}^2 - \kappa ( x \dot{y} - y \dot{x})^2 ] x 
+ \alpha^2 x =0,   \\
   \\
( 1 - \kappa r^2) \ddot{y} + \kappa [ \dot{x}^2 + \dot{y}^2 - \kappa ( x \dot{y} - y \dot{x})^2 ] y
+ \alpha^2 y =0.   \\
\end{array}
\label{eqII14}
\ba
Since the Schr\"odinger equation can be studied, it is worth remarking
about classical solutions.
The general solution to the Euler-Lagrange equations \eqref{eqII14}
will have the structure given as $(1)$ if $\kappa >0$ the dynamics is
restricted to the region $r^2 < 1/ |\kappa|$ where the kinetic energy is
positive definite and the general solution would be $x=  A \sin( \omega t + \phi_1)$,
$y = B \sin ( \omega t + \phi_2)$. $(ii)$ If $\kappa <0$ then the most general solution is
$x= A \sin ( \omega t + \phi_1)$, $y = B \sin ( \omega t + \phi_2)$ when the energy $E$
is smaller than a certain value $E_0$ and by $x = A \sinh ( \Omega t + \phi_1)$,
$y = B \sinh ( \Omega t + \phi_2)$ when the energy $E$ is greater than this value.
The coefficients $A$ and $B$ are related to $\alpha$ and the frequency $\omega$,
which is oscillatory motion,  or with
$\alpha$ and $\Omega$, unbounded motion.

\section{Calculation of Killing Vector Fields}

It is required to determine three linearly independent Killing vector
fields for the metric. For a vector field to be such a vector field, the 
Lie derivative of the metric \eqref{eqII2} with respect to this
vector field must vanish. A procedure will be presented 
which allows the determination of these vector fields from
the metric in closed form. Define a
general vector field $X$ with unknown coefficients which depend
on the cylindrical coordinates of the manifold as follows
\be
X = f ( r, \varphi) \frac{\partial}{\partial r} + h(r, \varphi) \frac{\partial}{\partial 
\varphi}.
\label{eqIII1}
\ba
The functions $f$ and $h$ are determined in such a way that the Lie
derivative of metric \eqref{eqII2} with respect to $X$ in \eqref{eqIII1} vanishes,
\be
{\cal L}_X g =0.
\label{eqIII2}
\ba
This is the condition that must be satisfied for $X$ to be a Killing vector field.
Using \eqref{eqIII1} in \eqref{eqIII2}, the Lie derivative of $g$ is found to be
$$
{\cal L}_{X}  g 
= X (\frac{1}{1 - \kappa r^2}) \, dr \otimes dr + \frac{1}{1- \kappa r^2} \{
\frac{\partial f}{\partial r} dr \otimes dr + \frac{\partial f}{\partial \varphi}
d \varphi \otimes dr + \frac{\partial f}{\partial r} dr \otimes dr
+ \frac{\partial f}{\partial \varphi} dr \otimes d \varphi \} + X (r^2) \,
d \varphi \otimes d \varphi
$$
\be
+ r^2 \{ \frac{\partial h}{\partial r} \, d r \otimes d \varphi
+ \frac{\partial h}{\partial \varphi} \, d \varphi \otimes d \varphi
+ \frac{\partial h}{\partial r} \, d \varphi \otimes d r
+ \frac{\partial h}{\partial \varphi} \, d \varphi \otimes d \varphi \}.
\label{eqIII3}
\ba
In order for \eqref{eqIII2} to hold, the coefficient of each tensor
product in \eqref{eqIII3} must vanish. Collecting like products, 
condition \eqref{eqIII2} is satisfied when the following system of partial differential 
equations is satisfied
\be
\frac{\partial f}{\partial r} + \frac{\kappa r}{1 - \kappa r^2} f =0,
\qquad
r^2 \frac{\partial h}{\partial r} + \frac{1}{1- \kappa r^2} 
\frac{\partial f}{\partial \varphi} =0,
\qquad
r \frac{\partial h}{\partial \varphi} =- f.
\label{eqIII4}
\ba
The general solution to system \eqref{eqIII4} is given as
\be
f (r, \varphi) = \sqrt{1 - \kappa r^2} ( c_1 \sin \varphi
+ c_2 \cos \varphi ),
\qquad
h (r, \varphi) = \frac{1}{r} \sqrt{1 - \kappa r^2} ( c_1 \cos \varphi
- c_2 \sin \varphi ) + c_3.
\label{eqIII5}
\ba
A set of three independent vector fields which satisfies \eqref{eqIII2}
will suffice. By picking three independent sets of constants $\{ c_i \}_{i=1}^3$
appropriately, these vector fields take the form 
\be
X_1 = \sqrt{1 - \kappa r^2} ( \cos \varphi \frac{\partial}{\partial r}
- \frac{1}{r} \sin \varphi \frac{\partial}{\partial \varphi}),
\qquad
X_2 = \sqrt{1 - \kappa r^2} ( \sin \varphi \frac{\partial}{\partial r}
+ \frac{1}{r} \cos \varphi \frac{\partial}{\partial \varphi}),
\qquad
X_J = \frac{\partial}{\partial \varphi}.
\label{eqIII6}
\ba
The set of vector fields \eqref{eqIII6} are the required Noether symmetries, and the coefficient
functions satisfy system \eqref{eqIII4}.
The associated constants of the motion are given by
\be
P_1 = \sqrt{1 - \kappa r^2} ( \cos \varphi \, p_r - \frac{1}{r}
\sin \varphi \, p_{\varphi} ),
\qquad
P_2 = \sqrt{1 - \kappa r^2} ( \sin \varphi \, p_r + \frac{1}{r}
\cos \varphi \, p_{\varphi}),
\qquad
J= p_{\varphi},
\label{eqIII7}
\ba
in the Hamiltonian formalism. The classical Poisson
bracket of any of the quantities $F,G$ from \eqref{eqIII7} is defined by
\be
\{ F, G \} = \frac{\partial F}{\partial r} \frac{\partial G}{\partial p_r}
+ \frac{\partial F}{\partial \varphi} \frac{\partial G}{\partial p_{\varphi}}
- \frac{\partial F}{\partial p_r} \frac{\partial G}{\partial r}
- \frac{\partial F}{\partial p_{\varphi}} \frac{\partial G}{\partial \varphi}.
\label{eqIII8}
\ba
Substitute the variables in \eqref{eqIII7} for $F$ and $G$ in
\eqref{eqIII8}, and the following brackets are obtained
\be
\{ P_1, P_2 \} = \kappa J,
\qquad
\{ P_1, J \} =- P_2,
\qquad
\{ P_2, J \} = P_1.
\label{eqIII9}
\ba
Moreover, using the Hamiltonian \eqref{eqII12} and \eqref{eqIII7}, the following
brackets can be calculated
\be
\{ P_1, H \} =0,
\qquad
\{ P_2 , H \} =0,
\qquad
\{ J, H \} =0.
\label{eqIII10}
\ba
The Lie brackets of the vector fields in \eqref{eqIII6} can also be
worked out. It is found that they close  in the following way,
\be
[ X_1, X_2 ] =- \kappa X_{J},
\qquad
[ X_1, X_J ] = X_2,
\qquad
[ X_2, X_J ] =- X_1.
\label{eqIII11}
\ba
Depending on the sign of $\kappa$, the Lie algebra of the group  of
isometries of the spherical, Euclidean and hyperbolic spaces is obtained. 
Only in the Euclidean case $\kappa =0$ do $X_1$ and $X_2$ commute.

It should be mentioned that although some of these calculations appear
tedious, they can be carried out very efficiently by means of symbolic
manipulation {\bf [15]}. Now \eqref{eqIII7} can be solved as a system for the
variables $ p_r, p_{\varphi}$, which are then placed into \eqref{eqII12} for the
Hamiltonian. The Hamiltonian then assumes the form,
\be
H ( \kappa) = \frac{1}{2m} [ P_1^2 + P_2^2 + \kappa J^2] - V(r).
\label{eqIII12}
\ba
This Hamiltonian will be used in the quantization procedure. A factor of
mass $m$ has been included to that end.
The only measure on the space $\mathbb R^2$ that is invariant
under the action of the vector fields \eqref{eqIII6} in the
sense that the Lie derivative vanishes,
$$
{\cal L}_{X_i} \,
d \mu_{\kappa} =0,  \qquad i=1,2,J
$$ 
is given up to a constant factor by
\be
d \mu_{\kappa} = \frac{r}{\sqrt{1 - \kappa r^2}} \, dr \wedge \, d \varphi.
\label{eqIII13}
\ba
To verify \eqref{eqIII13}, the Lie derivative of the right-hand side 
is evaluated with respect to each vector field in \eqref{eqIII6}
using the usual rules for Lie differentiation.

\section{Quantization and Schr\"odinger Equation}

This property of the measure suggests a way to quantize the Hamiltonian
for the model {\bf [12-13]}. The idea is to consider functions and linear operators
which are defined on a related space. This is obtained by taking the
two-dimensional real plane $\mathbb R^2$ and using the measure \eqref{eqIII13}
on it. To put it another way, the operators $\hat{P}_1$ and $\hat{P}_2$,
which represent the quantum version of the Noether momenta in \eqref{eqIII7}
must be self-adjoint not in the space $L^2 (\mathbb R)$, but in the space
$L^2 ( \mathbb R, d \mu_{\kappa})$ which is endowed with the measure 
\eqref{eqIII13}. 

Assuming the usual correspondence for the momenta in \eqref{eqIII7}, the
quantum operators which will form the quantum Hamiltonian are given by
$$
P_1 \raro \hat{P}_1 =- i \hbar \sqrt{1 - \kappa r^2}
[ \cos \varphi \frac{\partial}{\partial r} - \frac{1}{r} \sin \varphi
\frac{\partial}{\partial \varphi} ],
$$
\be
P_2 \raro \hat{P}_2 =- i \hbar \sqrt{1 - \kappa r^2} 
[ \sin \varphi \frac{\partial}{\partial r} + \frac{1}{r} \cos \varphi
\frac{\partial}{\partial \varphi} ],
\label{eqIV1}
\ba
$$
J \raro \hat{J} =- i \hbar \frac{\partial}{\partial \varphi}.
$$
Transforming \eqref{eqIII12} by means of \eqref{eqIV1}, the quantum Hamiltonian
is found to be
\be
\hat{H} = - \frac{\hbar^2}{2m} [ (1 - \kappa r^2) \frac{\partial^2}{\partial r^2}
+ (1- 2 \kappa r^2) \frac{1}{r} \frac{\partial}{\partial r} + \frac{1}{r^2}
\frac{\partial^2}{\partial \varphi^2}]
+ \frac{1}{2} \alpha^2 \frac{r^2}{1 - \kappa r^2}.
\label{eqIV2}
\ba
Using the operator \eqref{eqIV2}, it is straightforward to form the Schr\"odinger equation,
$\hat{H} \Psi = E \Psi$, as follows
\be
- \frac{\hbar}{2m} \{ (1- \kappa r^2) \frac{\partial^2}{\partial r^2}
+ (1 -2 \kappa r^2) \frac{1}{r} \frac{\partial}{\partial r}
+ \frac{1}{r^2} \frac{\partial^2}{\partial \varphi^2} \} \Psi
+ \frac{1}{2} \alpha^2 \frac{r^2}{1 - \kappa r^2} \Psi = E \Psi.
\label{eqIV3}
\ba
Before studying a class of solutions to \eqref{eqIV3}, it is helpful
to scale out the physical constants from equation \eqref{eqIV3}. To do this,
let us substitute the new variables $\bar{r}$, $\bar{\kappa}$ and ${\cal E}$
into the equation. 
To this end, let us set $\alpha = \sqrt{m} \beta$
and then make the replacement $\beta^2 \raro \beta^2 - (\kappa \hbar \beta)/m$
in the equation. The reason for the latter transformation is simply
that an $r$-dependent term can be factored from the equation thereby
simplifying it. Define
\be
r = \sqrt{\frac{\hbar}{m \beta}} \bar{r},
\qquad
\kappa = \frac{m \beta}{\hbar} \bar{\kappa},
\qquad
E = ( \hbar \beta) {\cal E}.
\label{eqIV4}
\ba
All of the physical constants in \eqref{eqIV3} simplify and factor out.
The Schr\"odinger equation takes the equivalent form,
\be
\{ ( 1 - \kappa r^2) \frac{\partial^2}{\partial r^2}
+ ( 1 - 2 \kappa r^2) \frac{1}{r} \frac{\partial}{\partial r}
+ \frac{1}{r^2} \frac{\partial^2}{\partial \varphi^2}
-(1- \kappa) \frac{r^2}{1 - \kappa r^2} \} \Psi =-2 {\cal E} \Psi.
\label{eqIV5}
\ba
In writing \eqref{eqIV5}, all the bars from \eqref{eqIV4}
have subsequently been dropped for ease of writing.

\section{Spectrum and Wavefunctions}

It is well worth studying Schr\"odinger equation \eqref{eqIV5} in this
case since many things can be learned with regard to the nature of
the spectrum and eigenfunctions of the equation. First of all, \eqref{eqIV5}
is a separable equation, and there exist solutions to it of the form
\be
\Psi (r, \varphi) = R(r) \Phi (\varphi),
\label{eqV1}
\ba
where $R$ and $\Phi$ are functions of the variables $r$ and $\varphi$.
Substituting $\Psi$ from \eqref{eqV1} into \eqref{eqIV5}, the equation becomes
\be
\Phi \{ (1 - \kappa r^2) R'' + (1 - 2 \kappa r^2 ) \frac{R'}{r} \}
+ \frac{1}{r^2} R \ddot{\Phi} - (1- \kappa) ( \frac{r^2}{1- \kappa r^2})
R \Phi =-2 {\cal E} R \Phi.
\label{eqV2}
\ba
Introducing a separation constant $\mu$, this can be written in the
separated form
\be
\frac{r^2}{R} \{ (1 - \kappa r^2) R'' + (1- 2 \kappa r^2) \frac{R'}{r} \}
- (1- \kappa)( \frac{r^4}{1 - \kappa r^2}) + 2 {\cal E} r^2 =
- \frac{\ddot{\Phi}}{\Phi} = \mu^2.
\label{eqV3}
\ba
This is equivalent to the following pair of ordinary equations
\be
\begin{array}{c}
\ddot{\Phi} + \mu^2 \Phi = 0,   \\
     \\
r^2 (1 - \kappa r^2) R'' + (( 1 - 2 \kappa r^2) r R' - (1 - \kappa)
\dss\frac{r^4}{1- \kappa r^2} R +2 {\cal  E} r^2 R - \mu^2 R =0,  \\
\end{array}
\label{eqV4}
\ba
The equation in $\Phi$ is easy to solve and 
has exponential solutions $e^{\pm i \mu \varphi}$.
The parameter $\kappa$ appears only in the radial equation in
the end.

The radial equation is the most work to solve, so consider it.
This equation has a factorized solution of the form
\be
R (r, \kappa) = F (r, \kappa) (1 - \kappa r^2)^s.
\label{eqV5}
\ba
Substituting \eqref{eqV5} into \eqref{eqV4}, the resulting
equation in terms of the parameter $s$ is given by
$$
r^2 ( 1 - \kappa r^2)^2 F'' - (1- \kappa r^2)(2 (2 s +1) \kappa r^2 -1) r F' 
+ (( 4 \kappa^2 s^2 + 2 \kappa^2 s + \kappa -1 -2 \kappa {\cal E} ) r^4 
$$
\be
+ ( \mu^2 \kappa - 4  \kappa \,s + 2 {\cal E}) r^2 - \mu^2) F =0.
\label{eqV6}
\ba
If $s$ is taken to be 
$$
s = \frac{1}{2} - \frac{1}{2 \kappa},
$$
the $r^4$ term disappears from the coefficient of the last term.
Moreover, evaluating the limit $\kappa \raro \infty$ with this $s$ at
fixed $r$ in \eqref{eqV5}, it is found to exist and is given by
$$
\lim_{\kappa \rightarrow 0} \, R (r, \kappa) = F(r) e^{r^2 /2}.
$$
Finally, equation \eqref{eqV6} simplifies to the form,
\be
r^2 (1 - \kappa r^2) F''  + ( 2 ( 1 - 2 \kappa) r^2 +1) r F' 
+ (2 (1 - \kappa +{\cal E})r^2 - \mu^2 ) F =0.
\label{eqV7}
\ba
This equation can be studied by means of the method of Frobenius and the
indicial equation for \eqref{eqV7} implies that it has a regular 
solution of the form
\be
F(r) = r^{\mu} \cdot f(r).
\label{eqV8}
\ba
The function $f(r)$ is regular at $r=0$. Substituting \eqref{eqV8}
into \eqref{eqV7}, it is found that $f(r)$ satisfies the equation
\be
r ( 1 - \kappa r^2) f'' + ( 2 ( 1- \kappa \mu - 2\kappa ) r^2 
+ 2 \mu +1) f' + ( 2 {\cal E} + 2 \mu - ( \mu +1)(  \mu +2) \kappa
 +2) r f =0.
\label{eqV9}
\ba
Assume now a $\kappa$-dependent power series for the function $f$
in \eqref{eqV9} of the form
\be
f (r, \kappa ) = \sum_{n=0}^{\infty} \, a_n (\kappa) r^n
\label{eqV10}
\ba
and substitute \eqref{eqV10} into \eqref{eqV9}. The following $\kappa$-dependent
recursion relation is obtained for the coefficients $a_n (\kappa)$,
$$
a_{n+1} ( \kappa) = \frac{(n + \mu)( n +\mu +1) \kappa -2 ({\cal E}
+ \mu +m)}{(n+1)(n+1 +2 \mu)} a_{n-1} (\kappa),
\qquad
n=1,2 , \cdots
$$
with $a_1 (\kappa)=0$ and $a_0 (\kappa)$ is an arbitrary constant.
The radius of convergence of the series \eqref{eqV10} is given by
$r_c = 1/ \sqrt{| \kappa|}$.

The even powers dependence implied by the recursion relation suggests
the introduction of the new variable $t= r^2$
\be
t (1 - \kappa t) f_{tt} + (1 + \mu + ( 1 - \kappa \mu - \frac{5}{2} \kappa)t ) f_t
+ \frac{1}{4} ( 2 {\cal E} + 2 ( \mu +1) - \kappa (\mu +1)(\mu +2)) f =0.
\label{eqV11}
\ba
When $\kappa =0$, which is the Euclidean case, the equation reduces to
\be
t f'' (t) + [ \mu + 1 +t] f' (t) + \frac{1}{2} ( {\cal E} + \mu +1) f(t) =0.
\label{eqV12}
\ba
There exists a solution which is regular at $r=0$ given by
\be
f(r) = c_0 e^{- r^2} K_M ( a , \mu+1 ; r^2 ),
\qquad
a = \frac{1}{2} (1 + \mu - {\cal E}).
\label{eqV13}
\ba
In \eqref{eqV13}, the function $K_M$ is the Kummer M-function. The 
physically acceptable solutions are the polynomial solutions which
appear when the parameter $a$ takes the values $a=- n_r$, $n_r=0,1,2, \cdots$.

When $\kappa \neq 0$, introduce the variable $s = \kappa t$ so the
equation is transformed into
\be
s (1-s) f_{ss} +(1 + \mu + \frac{2 - 2 \kappa \mu - 5 \mu}{2 \kappa} s) f_s
+ \frac{1}{4 \kappa} (2 {\cal E} + 2 \mu + 2 -  (\mu +1)(\mu +2) \kappa ) f =0.
\label{eqV14}
\ba
This is the Gauss hypergeometric equation in the standard form
\be
s (1 -s) f_{ss} +[ c - (1+ a_{\kappa} + b_{\kappa} )s ] f_s - a_{\kappa} b_{\kappa} f=0.
\label{eqV15}
\ba
The constants in \eqref{eqV15} are given by comparing with \eqref{eqV14},
\be
c= \mu +1 ,
\qquad
a_{\kappa} + b_{\kappa} = \frac{(2 \mu +3) \kappa -2}{ 2 \kappa},
\qquad
a_{\kappa} b_{\kappa} =- \frac{1}{2 \kappa} ({\cal E} + \mu +1)
+ \frac{1}{4} ( \mu +1)( \mu +2).
\label{eqV16}
\ba
The solution to \eqref{eqV14} is given by the hypergeometric function
\be
f (t) = \, _2F_1 ( a_{\kappa}, b_{\kappa} ; c; t),
\label{eqV17}
\ba
where $a_{\kappa}$ and $b_{\kappa}$ in \eqref{eqV17} are 
found to be given by
\be
a_{\kappa} = \frac{3 \kappa +2 \mu \kappa -2 - \Delta}{4 \kappa},
\qquad
b_{\kappa} = \frac{3 \kappa +2 \mu \kappa -2 +\Delta}{4 \kappa},
\qquad
\Delta= \sqrt{(\kappa -2)^2 + 8 {\cal E} \kappa}.
\label{eqV18}
\ba
The physically acceptable solutions which are determined as eigenfunctions
of the singular $\kappa$-dependent Sturm-Liouville problem appear when
one of the two $\kappa$-dependent coefficients $a_{\kappa}$, $b_{\kappa}$
coincides with zero or a negative integer number
\be
a_{\kappa} =- N_r,
\qquad
b_{\kappa} =-N_r,
\qquad
N_r =0,1,2, \cdots.
\label{eqV19}
\ba
This restricts the energy to one of the following values
\be
{\cal E} = (2 N_r + \mu +1) ( \frac{1}{2} ( 2 N_r + \mu +2) \kappa -1).
\label{eqV20}
\ba
The hypergeometric series then reduces to a polynomial of degree $N_r$.
Introducing the new quantum number $n = 2 N_r + \mu$, 
the energy levels are given by
$$
{\cal E} =  (n+1) (\frac{1}{2}(n+2) \kappa -1).
$$
Summarizing what has been obtained for this case, the wavefunctions
for the system on a space with constant curvature are given by
\be
\Psi_{N_r, \mu} (r, \varphi, ;\kappa) = C_{\kappa}
r^{\mu} (1 - \kappa r^2)^{\frac{1}{2} - \frac{1}{2 \kappa}} \,
_2F_1 ( - N_r, b_r; \mu+1; \kappa r^2) \, e^{\pm i \mu \varphi},
\label{eqV21}
\ba
where $C_{\kappa}$ is a normalization constant. 
Inverting transformation \eqref{eqIV4}, the energies 
for the model are then given by
\be
E_n (\kappa) = \frac{\hbar \alpha}{\sqrt{m}} (n+1) ( \frac{1}{2} 
( n+2) \kappa -1).
\label{eqV22}
\ba
The energy is a linear function of the curvature parameter $\kappa$
and it depends on the combination of the two quantum numbers
$2 N_r + \mu$. Finally, $\Psi_{N_r,\mu}$ is well defined for both
$\kappa >0$ and $\kappa <0$, and the degeneracy of the energy
levels is the same as in the Euclidean case.

\section{Concluding Observations}

It is worth summarizing some of the physical consequences
of what has been found here. The consequence of taking the metric
and potential in the form chosen is that the spectrum and
wavefunctions can be obtained for the system. The spherical case
corresponds to the parameter $\kappa$ in the interval $\kappa >0$.
The quantum Hamiltonian describes a quantum motion of a mass $m$
on a sphere $S_{\kappa}^2$. The particle possesses a countable
infinite set of bound states $\Psi_{N_r, \mu}$ and the energy
spectrum is unbounded, not equidistant and possesses a gap between
every consecutive pair that increases with $n$. The values are higher 
than in the Euclidean case.

The other case which has to be mentioned is the hyperbolic case 
$\kappa <0$. The Hamiltonian describes quantum dynamics on the hyperbolic
plane $H_{\kappa}^2$. In order for the wave function to be normalizable
with respect to the measure $d \mu_{\kappa}$, the limiting behavior 
of the square of the wave function times the $r$-dependent factor
in the measure enforces a limit on the number of bound states
in this case. The energy spectrum then becomes bounded not
equidistant, with a gap between every two levels that decreases.
In the hyperbolic case, the energies are lower than in the
Euclidean case as well.

\section{References}

\noindent
$[1]$ L. D. Landau and E. M. Lifshitz, Quantum Mechanics, (Pergamon Press Ltd,
Oxford, 1977).   \\
$[2]$ E. Prugovecki, Quantum Mechanics in Hilbert Space, (Academic Press, New York,
1971).   \\
$[3]$ N. D. Birrell and P. C. Davies, Quantum Fields in Curved Space,
(Cambridge University Press, Cambridge, England, 1994).   \\
$[4]$ L. E. Parker and D. J. Toms, Quantum Field Theory in Curved
Spacetime, (Cambridge University Press, Cambridge, England, 2009).     \\
$[5]$ P. Bracken, Hamiltonians for the Quantum Hall Effect on
Spaces with Non-Constant Metric, Int. J. Theoretical Phys.,
{\bf 46}, 119-132, (2007).   \\
$[6]$ E. Schr\"odinger, Proc. Roy. Irish Acad. Sect. {\bf A 46}, 9 (1940).  \\
$[7]$ P. W. Higgs, Dynamical Symmetries in a Spherical Geometry,
J. Phys. {\bf A 12}, 309-323, (1979).    \\
$[8]$ M. Lekshmanan and S. Rajackar, Nonlinear Dynamics, Integrability, Chaos
and Patterns (Springer-Verlag, Berlin, 2003).   \\
$[9]$ N. Woodhouse, Geometric Quantization, (Oxford University Press, Oxford, 1980).  \\
$[10]$ J. Sniatycki, Geometric Quantization and Quantum Mechanics 
(Springer, New York, 1980).   \\
$[11]$ I. H. McKenna and K. K. Wan, The role of the connection in geometric
quantization, J. Math. Phys. {\bf 25}, 1798-1803, (1984).   \\
$[12]$ J. F. Cari\~nena, M. F. Re\~nada and M. Santander, The quantum free 
particle on spherical and hyperbolic spaces: A curvature dependent approach,
J. Math. Phys. {\bf 52}, 072104, (2011).   \\
$[13]$ J. F. Cari\~nena, M. F. Ra\~nada and M. Santander,
The quantum Harmonic oscillator on the sphere and the hyperbolic plane:
$\kappa$-dependent formalism, polar coordinates and hypergeomrtic
functions, J. Math. Phys. {\bf 48}, 102106, (2007).  \\
$[14]$ J. F. Cari\~nera, M. F. Ra\~nada and M. Santander, 
A quantum exactly solvable non-linear oscillator with quasi-harmonic behavior,
Ann. Phys., {\bf 322}, 434-459, (2007).   \\
$[15]$ B. W. Char, K. O. Geddes, G. H. Gonnet, B. L. Leong, M. B. Monagen
and S. M. Watt, Maple V Library Reference Manual (Springer, New York, 1991).  \\
$[16]$ J. E. Marsden and T. S. Ratiu, Introduction to Mechanics and Symmetry
(Springer-Verlag, New York, 1994).  \\
\end{document}